\documentclass[aps,pra,reprint,groupedaddress,showpacs]{revtex4-1}

\usepackage{amsmath}
\usepackage[pdftex]{graphicx}

\usepackage[utf8]{inputenc}

\usepackage{pgfplots}

\usepackage{color} 
\usepackage{hyperref}

\newcommand{\fig}[1]{Fig.~\ref{#1}}
\newcommand{\eq}[1]{Eq.~(\ref{#1})}

\newcommand{\refsec}[1]{Sec.~\ref{#1}}

\newcommand{\BZ}{1BZ}
\newcommand{\kqG}{{\bf k}+{\bf q}}
\newcommand{\enk}[2]{\epsilon^{#1}_{#2}}

\newcommand{\etal}{\textit{et al.}}

\begin{document}

\normalsize

\title{Accurate bare susceptibilities from full-potential {\em ab initio} calculations}

\author{Christoph Heil}
\email[]{cheil@sbox.tugraz.at}
\affiliation{Institute of Theoretical and Computational Physics, University of 
Technology Graz, 8010 Graz, Austria}

\author{Heinrich Sormann}
\affiliation{Institute of Theoretical and Computational Physics, University of 
Technology Graz, 8010 Graz, Austria}

\author{Lilia Boeri}
\affiliation{Institute of Theoretical and Computational Physics, University of 
Technology Graz, 8010 Graz, Austria}

\author{Markus Aichhorn}
\affiliation{Institute of Theoretical and Computational Physics, University of 
Technology Graz, 8010 Graz, Austria}

\author{Wolfgang von der Linden}
\affiliation{Institute of Theoretical and Computational Physics, University of 
Technology Graz, 8010 Graz, Austria}

\date{\today}

\begin{abstract}
Electronic susceptibilities are a very popular tool to  study  electronic and magnetic properties of materials, both in experiment and theory.
Unfortunately, the numerical evaluation of even the bare susceptibility, which depends on the computation of matrix elements and sums over energy bands, is very work intensive and therefore various approximations have been introduced to speed up such calculations. We present a reliable and efficient implementation of the tetrahedron method which allows us to accurately calculate both static and dynamic bare susceptibilities, based on full-potential density functional theory (DFT) calculations. In the light of the exact results we assess the effects of replacing the matrix elements by a constant and the impact of truncating the sum over the energy bands. Results will be given for representative and topical materials such as Cr, a classical transition metal, as well as the iron-based superconductor FeSe.
\end{abstract}



\maketitle

\section{Introduction}
\label{sec:introduction}

The susceptibility $\chi$ both of charge and spin is a favoured quantity of theorists as it reveals interesting details of the excitation spectrum of the considered system and can directly be compared with experimental results~\cite{callaway_transverse_1975-1,charlesworth_susceptibilities_1995}.
There is a long list of materials for which susceptibilities have given important insights to unravel the underlying physics. In particular, for the iron-based superconductors~\cite{kamihara_iron-based_2008,johnston_puzzle_2010} (FeSCs), susceptibilities play a crucial role in the discussion of the origin of superconductivity. 

Many undoped FeSCs have an antiferromagnetic ground state, whose magnetic ordering vector matches the Fermi surface nesting vector~\cite{johnston_puzzle_2010}.
Since the very beginning of the field, this has been considered a strong indication of spin-fluctuation mediated superconductivity and many theories have been proposed to explain superconductivity based on susceptibility calculations~\cite{mazin_unconventional_2008,kuroki_unconventional_2008,chubukov_magnetism_2008,mazin_pairing_2009,monni_static_2010,ikeda_phase_2010,hirschfeld_gap_2011,thomale_mechanism_2011,essenberger_paramagnons_2012}.
In these weak-coupling approaches, the full electronic structure is usually reduced to an effective model using analytical approximations for the relevant bands (\textit{ab initio} downfolding or projection techniques) and the many-body interactions responsible for superconductivity, magnetism and other instabilities are treated with more and more sophisticated methods, such as random-phase approximation (RPA), fluctuation exchange approximation, functional renormalization group, etc.~\cite{ikeda_phase_2010,hirschfeld_gap_2011,thomale_mechanism_2011}.
Only very recently, a first-principles scheme for an \textit{ab initio} treatment of spin fluctuations has been proposed~\cite{essenberger_paramagnons_2012}.
Although they don't allow quantitative predictions of critical temperatures and energy scales, these calculations have provided very important insights into issues such as the symmetry of the order parameter, trends of superconducting critical temperatures, competition of different instabilities, and so on.

These studies have shown that even small changes in the electronic structure can have a large impact on superconductivity and magnetism. This implies that the influence of the approximations employed for the calculations of the non interacting system is a serious issue.
The non interacting susceptibility, representing the basis of RPA and of self-consistent DFT for spin fluctuations, is a particularly critical quantity:
the results are very sensitive to small details of the electronic structure, therefore very accurate electronic structures and $k$-space integration methods are needed; matrix elements are easy to compute in a plane wave basis, but converge very slowly with basis size; in cases where bare susceptibilities are used as input for many-body calculations, the number of bands is also a serious issue.

A common procedure in this case is to downfold the full electronic structure onto an effective low-energy model, which reproduces the band structure
in the vicinity of the Fermi level. This truncation can have severe effects on the susceptibility, since the convergence with the number of bands is very slow. In fact, although susceptibility calculations have been performed for a long time and many different algorithms have been proposed for the numerical evaluation~\cite{rath_generalized_1975,gilat_tetrahedron_1975,macdonald_extensions_1979, charlesworth_susceptibilities_1995}, the number of fully first-principles calculations of susceptibilities is 
scarce~\cite{samolyuk_character_2009,yaresko_interplay_2009,monni_static_2010,ke_low-energy_2011,essenberger_paramagnons_2012}.

In this paper we present a method that enables one to avoid these approximations and yields reliable results for the static and dynamic bare susceptibility, based on full-potential DFT calculations~\cite{blaha_wien2k_2001} of the electronic structure and tetrahedron integration~\cite{jepson_electronic_1971}.
We introduce a non standard tetrahedron method that explicitly takes only the non vanishing contributions of the Brillouin zone into account and significantly reduces the number of required $k$ points.
The accuracy of the presented approach is controlled by the number of $k$ points in the first Brillouin zone, the number of reciprocal lattice vectors $G$  used in the expansion of the LDA wave functions in plane waves, and the number of electronic bands entering the susceptibility formula. We demonstrate that converged results can be reached with acceptable computational effort.

Based on exact results obtained by this method, we scrutinize the approximation where all matrix elements are replaced by unity - henceforth referred to as 
{\it constant matrix element approximation} (CMA)~\cite{rath_generalized_1975,gilat_tetrahedron_1975,macdonald_extensions_1979,charlesworth_algebraic_1995}.
The CMA, which was very often used in the early days of susceptibility calculations, is at the heart of many qualitative arguments on Fermi surface nesting,
which have been revived in recent years to explain different phenomena, ranging from charge and spin density waves (SDWs) to superconductivity~\cite{mazin_unconventional_2008,borisenko_two_2009}.
We show that, in fact, CMA can strongly affect the susceptibility. Moreover, we discuss the effects of downfolding the full band structure to an effective low-energy model.

We would like to remark that the general purpose of this work is to provide a scheme to compute reliable bare susceptibilities. The inclusion of many-body effects, which would allow a direct comparison to experiment, is beyond the scope of this paper.

This paper is organized as follows.
In \refsec{sec:method} we introduce expressions for the bare
susceptibility $\chi^0$ and the method and algorithm we propose to
evaluate them with. In \refsec{sec:static_susceptibility} we study the
impact of  CMA on the static bare susceptibility $\chi^0$ of two
representative examples and we also discuss the error induced by
truncating the number of electronic bands. In
\refsec{sec:dynamic_susceptibility} we extend the analysis to dynamic
susceptibilities and our findings are summarized in
\refsec{sec:conclusions}.

\section{Method}
\label{sec:method}

\begin{figure*}
  \begin{center}
    \includegraphics[width=1.0\textwidth]{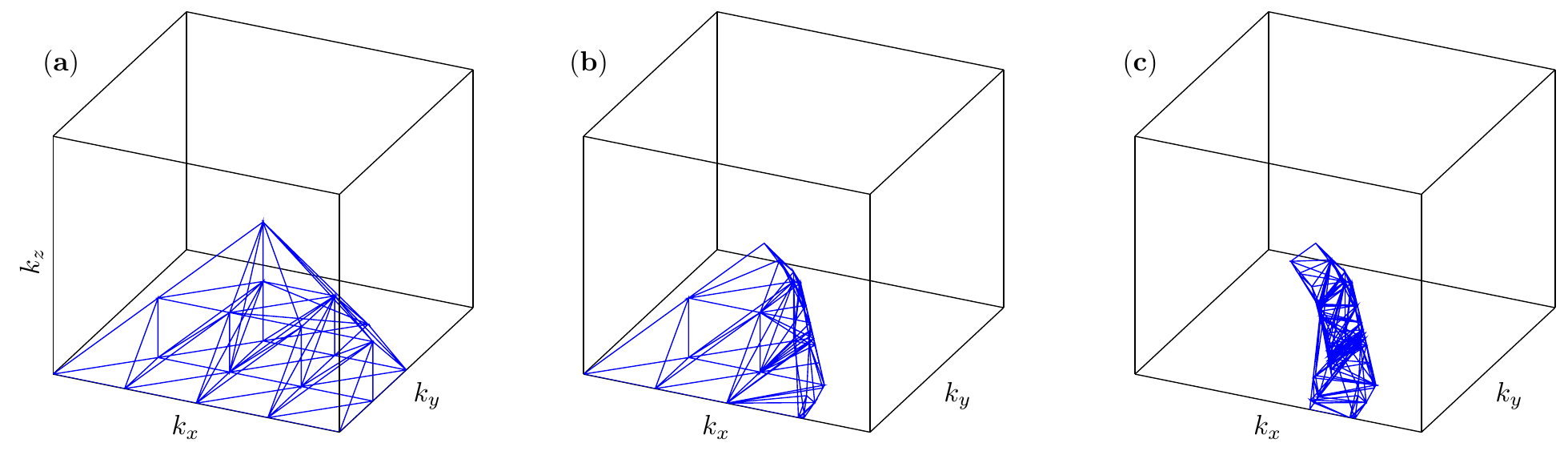}
  \caption{(Color online) $(\mathbf{a})$ Tetrahedral mesh of the irreducible wedge of an fcc lattice. 
  The other panels show the mesh after the first $(\mathbf{b})$ and
  after the second $(\mathbf{c})$ cut with the Fermi surface (see
  text). For the sake of simplicity we use a parabolic dispersion \mbox{$\epsilon(\mathbf{k}) = |\mathbf{k}|^2$; $\mathbf{q}= ( 0.15, 0, 0 )$ and $k_F=0.77$.}}
  \label{fig:tetrahedron_carving2}
  \end{center}
\end{figure*}
For a system of Bloch electrons - taking into account only the diagonal elements of the $\chi^0$ matrix - the real and imaginary parts of the non interacting ({\em bare}) dynamic susceptibility~\cite{fetter_quantum_2003} read
\begin{widetext}
\begin{align}
\mbox{Re}{\chi^0({\bf q},\omega)} &= \sum_{n,m}   
\int \limits^{\text{\BZ}}  \frac{d{\bf k}}{4\pi^3} \
\Theta(\epsilon_F-\enk{m}{\bf{k}}) 
\Theta(\enk{n}{\kqG}-\epsilon_F)
|\langle m,{\bf k}\vert {\rm e}^{-i{\bf q}\cdot{\bf r}}
\vert n,\kqG \rangle |^2
\left(
\frac{1}{\enk{m}{\bf k}-\enk{n}{\kqG}+\hbar\omega}
+
\frac{1}{\enk{m}{\bf k}-\enk{n}{\kqG}-\hbar\omega}
\right) 
\label{eq:chi0_2}\\
\text{Im}\chi^0({\bf q},\omega) &= -\sum_{n,m}   
\int \limits^{\text{\BZ}} \frac{d{\bf k}}{4\pi^2} \
\Theta(\epsilon_F-\enk{m}{\bf{k}}) 
\Theta(\enk{n}{\kqG}-\epsilon_F)
|\langle m,{\bf k}\vert {\rm e}^{-i{\bf q}\cdot{\bf r}}
\vert n,\kqG \rangle |^2 \ \delta(\hbar\omega+\enk{m}{\bf{k}}-\enk{n}{\kqG}),
\label{eq:chi0_3}
\end{align}
\end{widetext}
where \eq{eq:chi0_3} holds for $\omega \geq 0$. For negative values of $\omega$, the imaginary part is taken from
the relation 
\begin{align}
\text{Im}\chi^0({\bf q},-\omega) =  -\text{Im}\chi^0({\bf q},\omega)\,.
\label{eq:chi0_4}
\end{align}

The real and imaginary parts of $\chi^0$ are connected via the Kramers-Kronig transformation. ${\bf q}$ is a vector of the extended wave-vector space and ${\bf k}$ represents vectors of the first Brillouin zone (\BZ).
$n$ and $m$ denote electron band indices, $\epsilon_F$ is the Fermi energy, and $\enk{m}{\bf{k}}$ stands for the energy dispersion of the $m\text{th}$ band. The product of Heavi\-side functions in the numerator of the integrand ensures that only transitions from occupied to unoccupied electron states contribute to the integral. A similar expression for $\chi^0$ can be derived where the product of Heaviside functions is replaced by their difference $\Theta(\epsilon_F-\enk{m}{\bf{k}})-\Theta(\epsilon_F-\enk{n}{\kqG})$. While the latter is simpler from a geometrical point of view, it has the disadvantage that many contributions of the two terms cancel each other, an effect which becomes increasingly severe with decreasing $|\mathbf{q}|$.

We therefore chose Eqs.~(\ref{eq:chi0_2}) and (\ref{eq:chi0_3}) to be the basis of all our susceptibility calculations. In \refsec{sec:static_susceptibility}, which is dedicated to the {\em static} bare susceptibility, all results are obtained by an evaluation of \eq{eq:chi0_2} for $\omega=0$, in which case the imaginary part vanishes. 
However, if one wishes to study the {\em dynamics} of $\chi^0$ (as, e.g., in \refsec{sec:dynamic_susceptibility} of this paper),
one is usually interested in the $\omega$ dependence of both the real and the imaginary part of $\chi^0$. 
In this case, one normally computes only the imaginary part $\text{Im}\chi^0({\bf q},\omega)$ because its numerical evaluation is significantly less challenging than a direct calculation of $\text{Re}\chi^0({\bf q},\omega)$ and the corresponding real part can then be easily obtained by the Kramers-Kronig relation, provided that sufficiently large values of $\omega$ have been considered.

Both expressions (\ref{eq:chi0_2}) and (\ref{eq:chi0_3}) require a $k$-space integration over the irreducible wedge (IW) of the \BZ.
For a numerical evaluation of such integrals, different algorithms have been proposed in literature. 
Smearing methods are not appropriate for susceptibility calculations, and the most commonly used implementations are \textit{random sampling}~\cite{monni_static_2010,essenberger_paramagnons_2012} or tetrahedron methods~\cite{jepson_electronic_1971,lehmann_numerical_1972,rath_generalized_1975,gilat_tetrahedron_1975,blochl_improved_1994,charlesworth_algebraic_1995}.

In the following we present an implementation of the tetrahedron method that differs from other algorithms in some key aspects, which will be described later. The starting point, however, is the same as in all other implementations, i.e., the IW is decomposed into a number of tetrahedra as depicted in  \fig{fig:tetrahedron_carving2}(a) and described, for example, in Refs.~\onlinecite{lehmann_numerical_1972,gilat_tetrahedron_1975}.
The Bloch energies $\enk{m}{\bf{k}}$ and the corresponding wave functions $\psi_{m,{\bf k}}({\bf r})$ for $k$ points at the corners of the tetrahedra are determined by using electron band structure codes, in our case the {\em full potential linearized augmented plane wave} (FP-LAPW) code WIEN2k~\cite{blaha_wien2k_2001,andersen_linear_1975}.
Numerically, each wave function $\psi_{m,{\bf k}}({\bf r})$ is represented by a set of plane wave coefficients $a_{m,{\bf k}}({\bf K})$, with {\bf K} being the reciprocal lattice vector. Based on this system of input data, the energies $\enk{n}{\kqG}$ and coefficients  $a_{n,\kqG}({\bf K})$, which usually
belong to $k$ points {\em not} contained in the tetrahedra set, are approximated by linear interpolation. This means that in our calculations both
the electron energies and the matrix elements in Eqs.~(\ref{eq:chi0_2}) and (\ref{eq:chi0_3}) are numerically treated {\em on equal footing}.

The most demanding aspect of the numerical evaluation of the integrals (\ref{eq:chi0_2}) and (\ref{eq:chi0_3}) comes from the product of Heaviside functions
$\Theta(\epsilon_F-\enk{m}{\bf{k}})\Theta(\enk{n}{\kqG}-\epsilon_F)$: first, the step function $\Theta(\epsilon_F-\enk{m}{\bf{k}})$ reduces the integration within the IW to {\em initial} electron states $\epsilon_{\bf k}^m$ lying {\em inside} the Fermi surface (FS). The numerical realization of such a reduction is quite popular in the literature; some authors (see, e.g., Charlesworth and Yeung \cite{charlesworth_algebraic_1995}) call this a {\em geometric interpretation} of the tetrahedra to emphasize that their occupation depends on their spatial position within the IW: a tetrahedron is said to be either empty, fully occupied, or partially occupied, if it is either entirely outside, inside, or cut by the Fermi surface. Empty tetrahedra are removed as they do not contribute to the integral and fully occupied ones remain unchanged. Since the integration over fully occupied tetrahedra is less time consuming and more reliable than over partially occupied ones, the latter are further decomposed into 
a finer set of completely occupied tetrahedra.
This procedure, schematically shown in \fig{fig:tetrahedron_carving2}(a) and \fig{fig:tetrahedron_carving2}(b), is described in detail
by, e.g., MacDonald \etal~\cite{macdonald_extensions_1979} and Rath and Freeman~\cite{rath_generalized_1975}.

The key aspect of our implementation is that this process of cutting the tetrahedra to carve out the regions where the Heaviside function equals unity is consequently repeated also for $\Theta(\epsilon_{\kqG}^n-\epsilon_F)$. In this way, the \textit{final} tetrahedra are restricted to the region of the IW, where the condition
\begin{equation}
\Theta(\epsilon_F-\enk{m}{\bf{k}})
\Theta(\enk{n}{\kqG}-\epsilon_F)= 1\;
\label{eq:Theta1}
\end{equation}
is fulfilled.~\footnote{$\Theta(\enk{n}{\kqG}-\epsilon_F)$ reduces the integration within the IW to {\em final} electron states $\epsilon_{\bf k+\bf q}^n$ lying {\em outside} the Fermi surface.} The corresponding (second) reformulation of the set of tetrahedra in the IW is graphically demonstrated by 
\fig{fig:tetrahedron_carving2}(b) $\to$ \fig{fig:tetrahedron_carving2}(c). While the first step from (a) to (b) has to be performed only once for every electron band (of the sum over $m$), the second step has to be repeated (i) for every value of the ${\bf q}$ vector, (ii) for every electron band of the sum over $n$, and (iii) for every point group element of the crystal. The main benefit of this approach is that a lot of numerical issues encountered when integrating over tetrahedra, which do not fulfill \mbox{condition (\ref{eq:Theta1})} but only $\Theta(\epsilon_F-\enk{m}{\bf{k}})=1$, can be avoided and a simpler set of integration formulas can be used.
Besides the simplified numerical integration, the number of $k$ points in the IW can be significantly reduced. For example, the results for chromium, which we are going to discuss in \refsec{sec:static_susceptibility}, were obtained by using approximately $500$ $k$ points in the IW. To achieve the same accuracy without the second carving of the \BZ, we would have needed to consider more than $2000$ $k$ points.

Finally, we also tested a quadratic interpolation of the electron energies and transition matrix elements~\cite{blochl_improved_1994}, which requires DFT calculations for additional $k$ points. We found, however, that it is more advantageous to use these additional $k$ points directly to create a tighter tetrahedral mesh and to use a linear interpolation.

\section{Static Bare Susceptibility}
\label{sec:static_susceptibility}

The CMA is the simplest approximation to avoid the cumbersome evaluation of the matrix elements entering the bare susceptibility 
formula. This approximation has been used to interpret susceptibility data and SDW order in many transition metals, such as Pd and Cr. The concept of Fermi surface nesting, which derives from these early CMA calculations, is still quite popular nowadays in several materials, such as FeSCs and layered metal dichalchogenides. We will demonstrate that although CMA in some cases may lead to reasonable results, it fails in other cases. In the light of this unpredictability it is advisable to include matrix elements in all static susceptibility calculations and even more so in the dynamical ones.

In order to calculate the matrix elements $\langle m,\mathbf{k}| e^{-i\mathbf{q}\cdot\mathbf{r}} |n,\kqG\rangle$ correctly we expand the  LAPW~\cite{blaha_wien2k_2001} eigenvectors $|m,\mathbf{k} \rangle$ in a plane wave basis. We carefully checked that our results are converged with respect to the number of plane waves (typically around 4000).

We start our discussion with the 3$d$ transition metal chromium (Cr), a classical SDW material~\cite{shull_neutron_1953}, for which different 
approximations for susceptibilities have been proposed. Neutron scattering experiments show an incommensurate SDW  with $|\mathbf{q}_{\text{SDW}}| \sim \frac{21}{22} \frac{2 \pi}{a}$, which corresponds to one of the nesting vectors of the Fermi surface. Fermi surface nesting was however not sufficient to explain why Mo, which has a very similar Fermi surface, does not display any SDW. This led to more refined susceptibility calculations~\cite{windsor_interband_1972,fawcett_spin-density-wave_1988,schwartzman_concentration_1989},
which showed that the susceptibilities of the two materials with matrix elements are quite different, with no obvious SDW instability in Mo. 
\begin{figure}
  \begin{center}
    \includegraphics[width=1.0\linewidth]{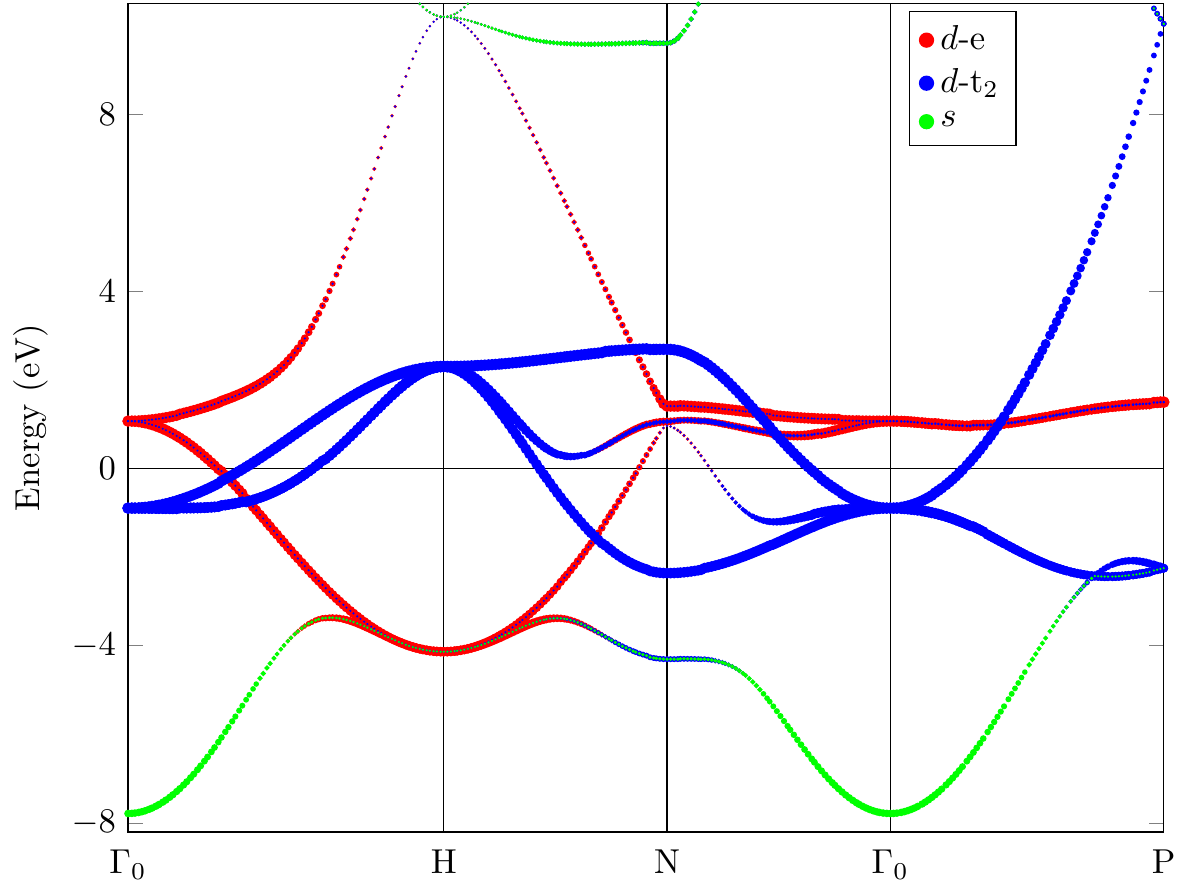}
  \caption{(Color online) LDA bandstructure of Cr, decorated with partial characters: $s$ (green), $d$-t$_2$ (blue), and $d$-e (red).
   The coordinates of the high-symmetry points are $\Gamma_0$=(0,0,0), H=(0,1,0), N=($\frac{1}{2}$,$\frac{1}{2}$,0), and P=($\frac{1}{2}$,$\frac{1}{2}$,$\frac{1}{2}$), all in units of $\frac{2 \pi}{a}$ with $a$ being the lattice constant.}
   \label{fig:Cr_bands}
   \end{center}
\end{figure}

The electronic structure of Cr in the experimental body centered cubic (BCC) crystal structure is shown in \fig{fig:Cr_bands}; the colored symbols indicate the partial character of the electronic bands: \mbox{$s$ (green)}, $d$-e, i.e.,  3$z^2-1$, $x^2-y^2$ (red) and $d$-t$_2$, i.e., $xy,xz$ and $yz$ (blue).
The $s$ band is entirely full, and extends from $\sim 8$ to $\sim 4$ eV below the Fermi level ($\epsilon_F$).
The $d$ bands lie higher, with a clear separation between t$_2$ bands, which form a narrow structure $\pm 2$\,eV around $\epsilon_F$, and e bands extending to higher energies.
In many cases, however, the bands are not {\em pure}, i.e., they display contributions from more than one partial character, and this indicates a substantial
hybridization between the corresponding real-space orbitals.
\begin{figure}
  \begin{center}
    \includegraphics[width=0.85\linewidth]{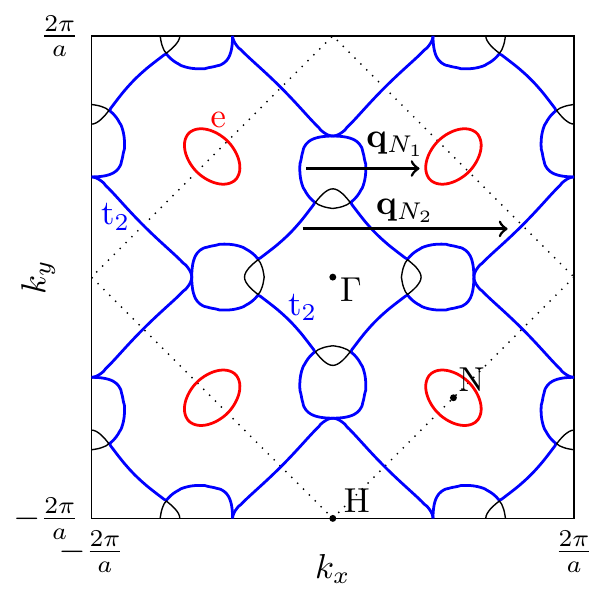}
  \caption{(Color online) FS of Cr in the $k_z$=0 plane. The different colors indicate different orbital character as described in \fig{fig:Cr_bands} and the boundary of the 1BZ is indicated by dotted lines.}
  \label{fig:Cr_FS}
  \end{center}
\end{figure}

The resulting Fermi surface is three dimensional and comprises three types of sheets: large octahedral hole and electron pockets around the center ($\Gamma$) and at the corner (H) of the BZ, and several smaller hole pockets around the N points.
A two-dimensional section of the FS in the $k_z$=0 plane is shown in \fig{fig:Cr_FS}; the color coding for the dominant partial characters is the same as in \fig{fig:Cr_bands}.
The large pockets around $\Gamma$ and H are mostly of t$_2$ character, the small  ellipsoids around N are mostly of e character.

\begin{figure}
  \begin{center}
    \includegraphics[width=1.0\linewidth]{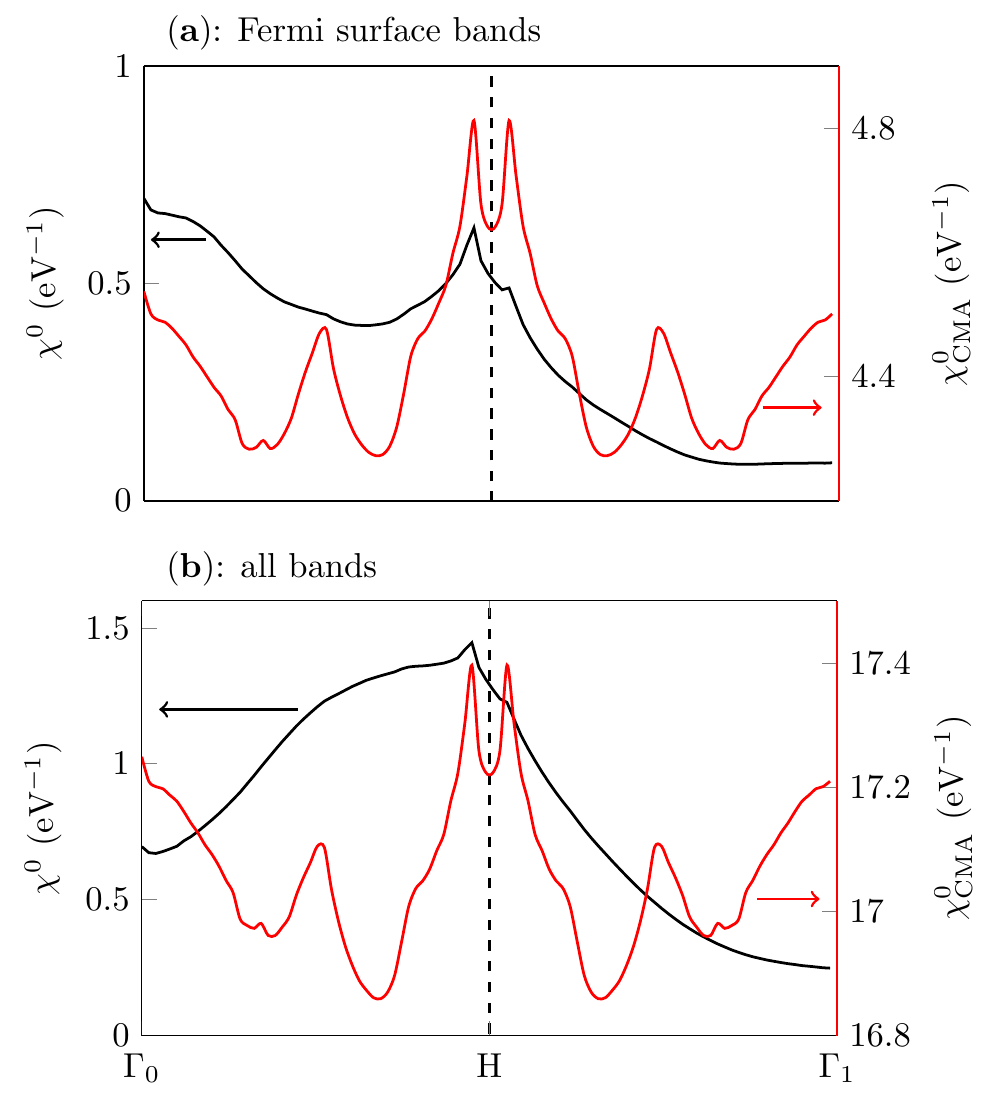}
  \caption{(Color online) Static bare susceptibility $\chi^0$ for Cr. CMA results (red, right  axis) are compared with the exact results (black, left axis). 
$\Gamma_{1}$ stands for the $\Gamma$ point in the second BZ.  
  The upper panel ($\mathbf{a}$) shows the converged results, while in the lower panel ($\mathbf{b}$)
only the contributions of the bands at the Fermi level have been included in the susceptibility.}
  \label{fig:Cr_rechi0_comp}
  \end{center}
\end{figure}

Based on this electronic structure, we have used our highly accurate susceptibility program to compute the corresponding $\chi^0$ along the $(010)$-direction from the $\Gamma$ point of the \BZ \ ($\Gamma_0$) to the center of the next BZ ($\Gamma_1$). In \fig{fig:Cr_rechi0_comp} we compare exact results where matrix elements are properly taken into account (black curve) with the CMA (red curve).
The left scale belongs to the exact results and the right scale to CMA. The cross at $|\mathbf{q}|=0$ marks the value of the density of states at the Fermi energy $N(\epsilon_F)$ given by the LAPW calculation, a value $\chi^0(\mathbf{q})$ has to approach in the limit $|\mathbf{q}| \rightarrow 0$. Of course, $N(\epsilon_F)$ can only be calculated reliably when matrix elements are correctly included. The agreement of $\lim_{\mathbf{q} \to 0} \chi_{0}(\mathbf{q})$ with the exact value of the density of states is a stringent test for the $k$-space integration.

The exact result yields a broad bell-shaped curve with a small and narrow peak at $\mathbf{q} =\mathbf{q}_{N_2}$, with 
\mbox{$|\mathbf{q}_{N_2}| \approx 0.95$\,$\frac{2\pi}{a}$}, 
which fits perfectly to the experimentally observed wavelength of the SDW in chromium,~\cite{shull_neutron_1953} and is in very good agreement with previous tight-binding~\cite{windsor_interband_1972} and supercell calculations.~\footnote{For a review of SDW in Cr, see M. Bayer, PhD Thesis, TU Dresden 2008}
Although there is a peak at the same wave number in the CMA result (red curve) as well, the rest of the susceptibility differs significantly from the exact result. In particular, we observe a second very strong peak at 
$\mathbf{q} =\mathbf{q}_{N_1}$, with 
\mbox{$|\mathbf{q}_{N_1}| \approx \frac{\pi}{a}$}, which is not present in the exact result. 

In order to understand these results in detail, we first compare the converged susceptibility [\fig{fig:Cr_rechi0_comp}(b)], which was obtained by summing over $30$ bands, with the result we would obtain restricting the sum in \eq{eq:chi0_2} only to bands at the Fermi level [\fig{fig:Cr_rechi0_comp}(a)]. 
First of all, we notice that the red curves in panels (a) and (b) are almost identical, i.e., in the CMA the shape of $\chi^0$ is almost entirely determined by the transitions between bands at the Fermi level. The situation is very different for the exact susceptibility (black curves), where matrix elements strongly enhance transitions between ``outer'' bands, i.e., bands that do not cross the Fermi level; in this particular case, this enhancement is very strong for $k$ points half-way between $\Gamma$ and H. 

While it is almost impossible to give a detailed account of all the transitions involving outer bands, since their number is very large, it is extremely instructive to trace back the enhancement
of the susceptibility due to matrix elements, when only bands at the Fermi surface are included [panel $\mathbf{a}$].
Indeed, the two peaks seen at q$_{N_1}$=$0.52$\,$\frac{2\pi}{a}$ and q$_{N_2}$=$0.95$\,$\frac{2\pi}{a}$ in the CMA correspond to two nesting vectors of the Fermi surface, illustrated in \fig{fig:Cr_FS}.
The shorter vector (q$_{N_1}$) connects the large hole FS with small ellipsoidal pockets around N, while the larger one (q$_{N_2}$) connects it to the large electron pocket at H.
The large difference in the matrix elements stems from the fact that q$_{N_2}$ connects parts of the Fermi surface for which not only the geometrical, but also the orbital overlap, is large, and this does not happen for q$_{N_1}$.~\footnote{The effect of orbital overlap is easy to understand in the tight-binding approximation,~\cite{windsor_interband_1972,kemper_sensitivity_2010} where the matrix elements in Eqs.~(\ref{eq:chi0_2}) and (\ref{eq:chi0_3}) are approximated by
$ 
\sum_{i}c_i^{n}(\mathbf{k})c_i^{m}(\mathbf{k+q})\,.
$ 
Here, $c_i^{n}(\mathbf{k})$ are the normalized coefficients of the $i^{\text{th}}$ atomic orbitals for the electronic Bloch state of band $n$ at wave-vector $\mathbf{k}$.}

The main conclusions of our first detailed comparison is that the CMA is a very poor approximation for the full susceptibility for two reasons: (i) it overemphasizes the role of the bands that cross the Fermi surface; (ii) it neglects completely the information on the orbital character of the electronic states, which has a major effect on matrix elements.
Finally, we also want to remark that in CMA, the $\mathbf{q}$-dependence of the susceptibility solely stems from the energies $\epsilon_{\mathbf{k}+\mathbf{q}}$, which are periodic with respect to any reciprocal lattice vector $\mathbf{K}$ of the crystal. Therefore one has $\chi^0(\mathbf{q})=\chi^0(\mathbf{q}+\mathbf{K})$, as observed in all red curves of Figs.~\ref{fig:Cr_rechi0_comp}, which is an artifact of the approximation.~\footnote{For a more detailed discussion of momentum dependence of the matrix elements see Refs.~\onlinecite{gupta_wave-number-dependent_1971,gupta_matrix_1976}.}

After the classical example of Cr, we next present results for FeSe as a representative example of the wide class of the recently-discovered FeSCs~\cite{kamihara_iron-based_2008,johnston_puzzle_2010}, where models based on susceptibilities have played a major role in the past few years.
For the following discussion we have chosen FeSe because it is one of the ``simplest'' FeSCs in terms of chemical formula and crystal structure. 
This  allows us to discuss the electronic structure without entering the details of hybridization with intercalated atoms and layers and three-dimensional interlayer hopping; to avoid the problem of unfolding, we also chose to work in the two-Fe unit cell.

\begin{figure}
  \begin{center}
    \includegraphics[width=1.0\linewidth]{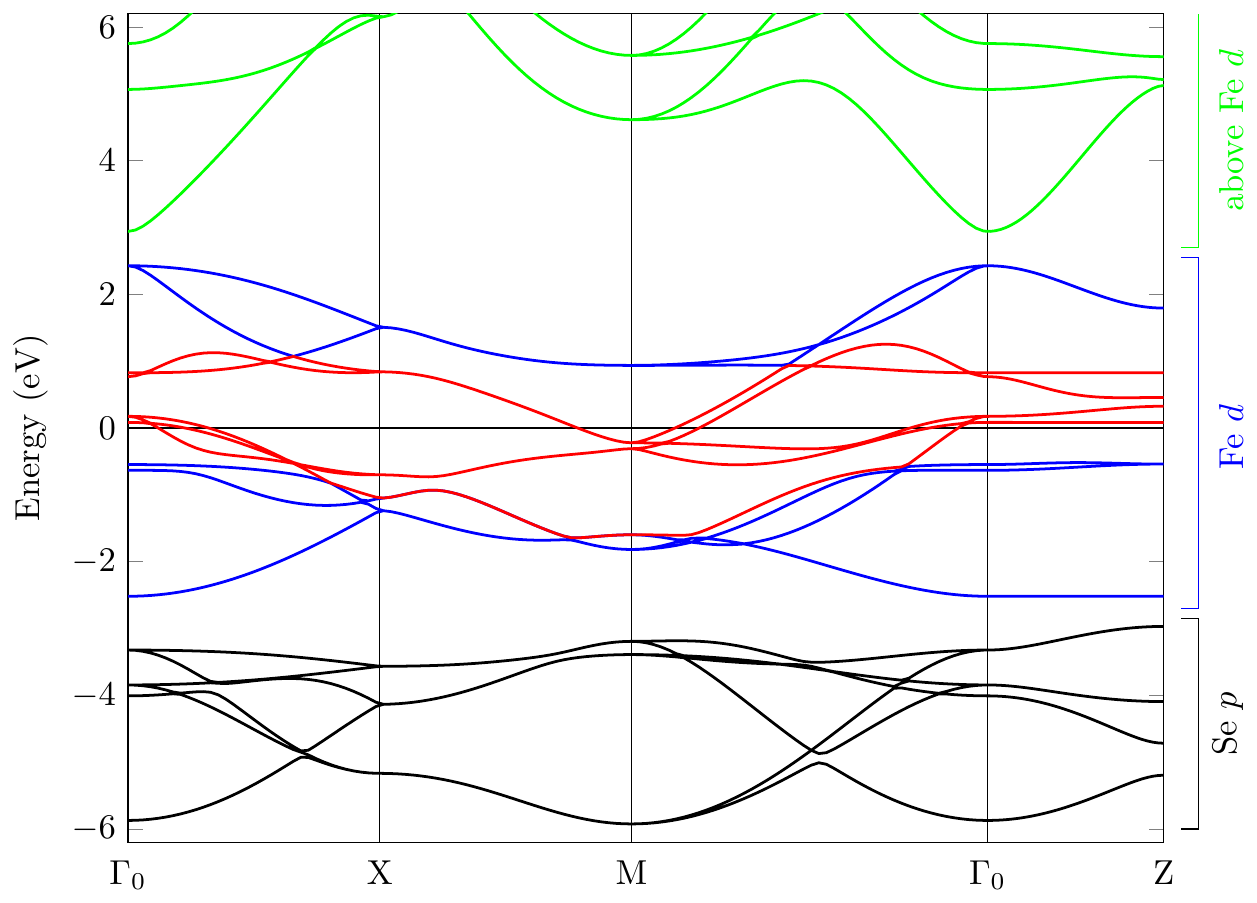}
  \caption{(Color online) LDA bandstructure for FeSe at ambient pressure with the bands labeled according to their dominant orbital character. The five Fe $d$ bands that form the Fermi surface are shown in red, while all other \mbox{Fe $d$} bands are depicted in blue. The \mbox{Se $p$} are shown in black and the bands above \mbox{Fe $d$} in green.}
  \label{fig:FeSe_0_bands}
  \end{center}
\end{figure}

All calculations presented here employ the crystal structure at ambient pressure measured by Kumar~\etal \ in Ref.~\onlinecite{kumar_crystal_2010}.
Our electronic structure, shown in \fig{fig:FeSe_0_bands}, agrees nicely with previous studies~\cite{subedi_density_2008,kumar_crystal_2010};
similarly to what we did in \fig{fig:Cr_bands}, we have colored the bands according to their dominant character and/or position with respect to the Fermi level.
This choice allowed us to introduce a compact notation for the susceptibility plots, but it does not permit one to appreciate the full complexity of the electronic structure. This issue is discussed in more detail in other publications~\cite{kuroki_pnictogen_2009,kemper_sensitivity_2010,andersen_multi-orbital_2011}. Here we only want to recall that, due to the sizable $p-d$ hybridization, there is a substantial contribution of Se $p$ states to the Fe $d$ bands, and vice versa.
 
The sixteen Fe $d$ - Se $p$ bands form a manyfold which extends from $\sim -6$ to $\sim + 2$\,eV around the Fermi energy; the six lowest bands have mostly selenium character, and are separated by a small gap from the ten Fe bands  at $\pm 2$\,eV. 
The Fermi level cuts the iron bands at a nominal electron count $d^6$, creating three hole pockets at the $\Gamma$ point, and two electron pockets at the M point. The $k_z$ dispersion of the bands is so small that the FS is essentially two dimensional.

The inner and outer hole pockets  have dominant $xz,yz$ orbital character, while the middle hole pocket is mostly of $xy$ character;
the electron pockets are formed by two ellipsoids with the long axis along the 110 and 1$\bar{1}$0 directions,
with dominant $xz/yz$ character on the long side and $xy$ on the short one.

A clear geometrical nesting for $\mathbf{q} \sim \text{M} =(\frac{\pi}{a},\frac{\pi}{a},0)$ exists between the hole and electron pockets; 
this feature is common to many Fe-based superconductors, but the different shape and orbital composition in different compounds can lead to marked differences in 
the full susceptibility. Note also that all  partial character of the hole and electron pockets match over a considerable part of the BZ.

\begin{figure}
  \begin{center}
    \includegraphics[width=1.0\linewidth]{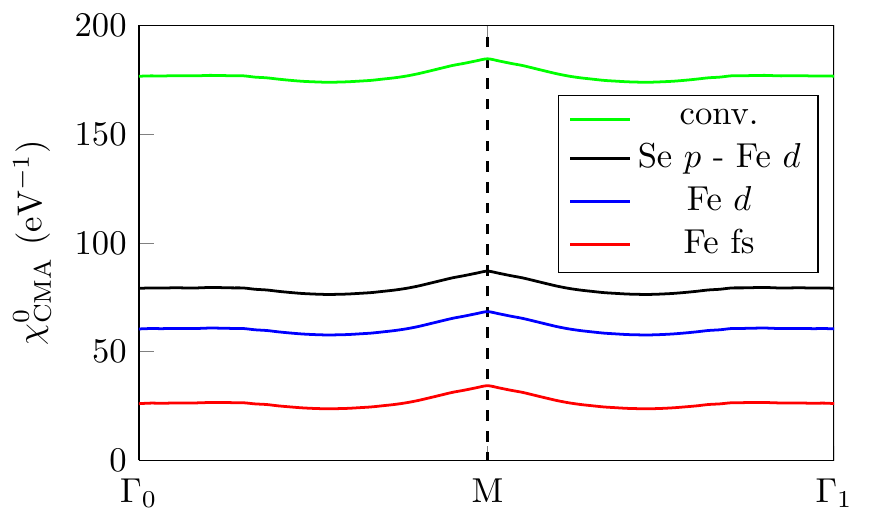}
  \caption{(Color online) Static bare susceptibility in  CMA $\chi^0_{\text{CMA}}$ for FeSe calculated by including different sets of bands, as shown in \fig{fig:FeSe_0_bands}. From bottom to top the included bands are: Fe $d$ that cross the Fermi energy (red), all Fe $d$ bands (blue), Se $p$ plus all Fe $d$ bands (black), and Se $p$ plus Fe $d$ plus all higher bands which are needed to ensure convergence of the full susceptibility $\chi^0$.}
  \label{fig:FeSe_0_rechi0_cm}
  \end{center}
\end{figure}

\begin{figure*}
  \begin{center}
    \includegraphics[width=0.9\linewidth]{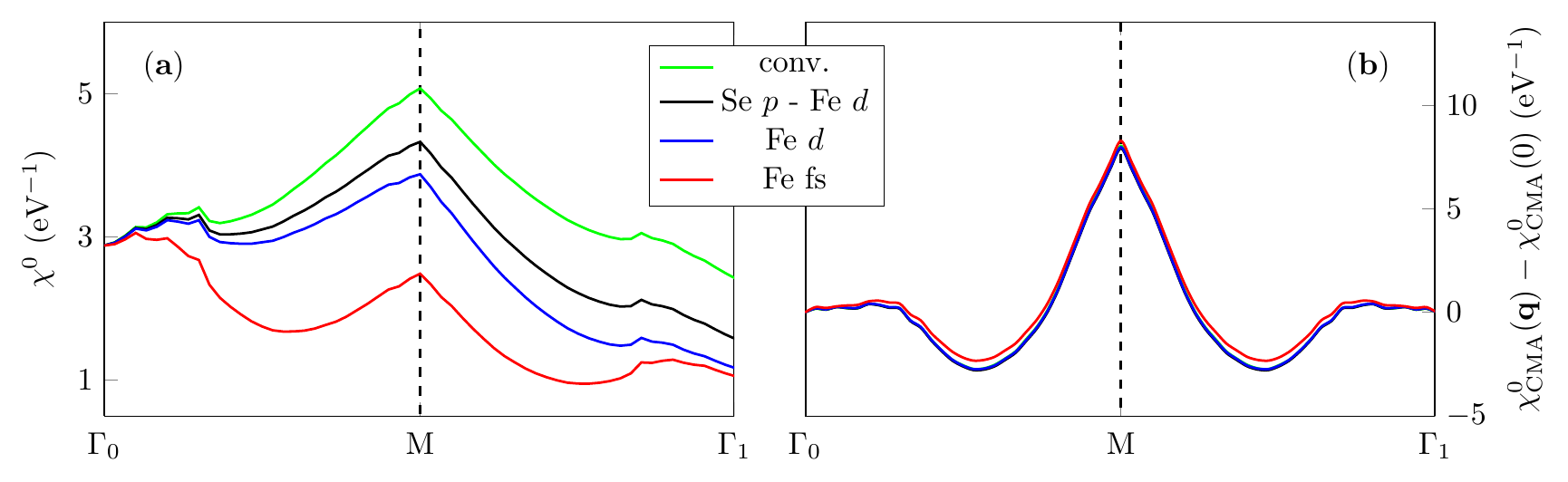}
  \caption{(Color online) Static bare susceptibility $\chi^0$ for FeSe at ambient pressure calculated by including different sets of bands, as explained in \fig{fig:FeSe_0_rechi0_cm}. The left panel $(\mathbf{a})$ shows the exact result while the right panel $(\mathbf{b})$ contains the CMA result.}
  \label{fig:FeSe_0_rechi0_comb}
  \end{center}
\end{figure*}
We analyze its behavior in detail for FeSe, studying $\chi^{0}$ in the (110)-direction in $k$-space. We start from the CMA picture, shown in \fig{fig:FeSe_0_rechi0_cm}. 
The color coding in the figure is consistent with \fig{fig:FeSe_0_bands}: $\chi^{0}_{\text{CMA}}$ results calculated considering only
transitions between the Fe bands that create the Fermi surface are drawn in red, those which also involve the rest of the \mbox{Fe $d$} bands not crossing $\epsilon_F$ in blue, while in black we have all transitions from \mbox{Se $p$} to \mbox{Fe $d$}. The converged results with respect to the number of bands is shown in green.

In $\chi^0_{\text{CMA}}$, since matrix elements are neglected and the denominator of \eq{eq:chi0_2} is almost $k$ independent for large energies, the inclusion of more bands in the sum results in an almost rigid shift in the susceptibility, which decreases as $1/\Delta \epsilon$ for bands away from the Fermi level.
This background shift has no physical meaning, and in order to compare susceptibility curves with different numbers of bands, it is more meaningful to shift them to a common offset. This is done in the right panel of \fig{fig:FeSe_0_rechi0_comb}, where $\chi^0_{\text{CMA}}$ is set to zero at $|\mathbf{q}|=0$ for all curves.
Due to the constant matrix elements, $\chi^0_{\text{CMA}}$ depends purely on  $1/(\enk{m}{\mathbf{k}}-\enk{n}{\kqG})$. This expression is large only for the partially filled Fermi surface bands and \mbox{$\chi^0_{\text{CMA}}(\mathbf{q}) - \chi^0_{\text{CMA}}(0)$} therefore depends mainly on these bands; the most evident feature is a pronounced peak at the M point, due to the nesting of hole and electron Fermi sheets. An enhancement is seen also around the $\Gamma$ point, for $\mathbf{q} \lesssim 0.4$\,$\overline{\Gamma \text{M}}$ due to hole-hole and electron-electron transitions; no inter- or intra-band transitions
are possible for $ 0.4  \lesssim |\mathbf{q}| \lesssim 0.6$\,$\overline{\Gamma \text{M}}$, and this accounts for the depletion seen in $\chi^0_{\text{CMA}}$ for these values of $|\mathbf{q}|$.

The full susceptibility, with matrix elements correctly taken into account, is shown in the left panel of \fig{fig:FeSe_0_rechi0_comb}. We want to stress that in this case no scaling or shifting of the results has been performed. As long as all bands which cross the Fermi level are included in the calculation, $\chi^0$ approaches $N(\epsilon_F)$ in the limit $|\mathbf{q}| \rightarrow 0$. Away from $\Gamma_0$ the absolute value does of course depend on the number of included bands. All curves have a  peak at the M point. However, quite surprisingly, the absolute maximum of  the red curve, calculated based only on the Fermi surface bands, is not at M, but close to $\Gamma_0$. This means that at the Fermi surface the matrix element enhances hole-hole and electron-electron transitions more than electron-hole ones.
Note that based on this result, we could conclude that this particular system has a dominant instability at small $|\mathbf{q}|$, at variance with most other FeSCs.
However, the full susceptibility, including bands  fairly  away from the Fermi energy, has its maximum at M; the convergence to the exact curve in terms of included bands is quite slow~\footnote{We found that the results for the full $\chi^0$ converge if approximately $50$ bands around $\epsilon_F$ are included.}.
\begin{figure}
  \begin{center}
    \includegraphics[width=0.6\linewidth]{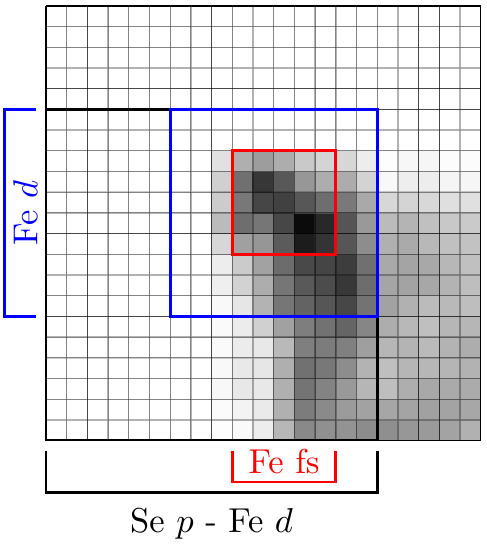}
  \caption{(Color online) Contributions of the band transitions to $\chi^{0}$ for FeSe 
at ambient pressure at the M point. The red square contains transitions only between Fermi surface bands and the blue square contains all band transitions between the Fe 3$d$ bands. Inside the black square are all \mbox{Fe $d$} and \mbox{Se $p$} bands. We employed a logarithmic color scale to visually enhance small values.}
  \label{fig:FeSe_0_rechi0_contributions}
  \end{center}
\end{figure}

This is also graphically illustrated in \fig{fig:FeSe_0_rechi0_contributions}, where the contributions of the individual band transitions to the susceptibility $\chi^0$ at the M point are represented in a two-dimensional histogram.  The red square contains all bands that create the Fermi surface, while the blue square includes all transitions between Fe $d$ bands and the black square all Fe $d$ and Se $p$ bands. The biggest contribution originates from the transition of the middle hole pocket to the outer electron pocket of the Fermi surface. One can also observe that there are considerable contributions to the susceptibility outside the red square and also outside the blue square, again substantiating  the vital importance of including enough bands in a $\chi^0$ calculation. We note in passing that similar calculations for other FeSCs (not shown) display a different convergence as a function of the number of included bands.

We can summarize this section noting that in systems like FeSCs,  with a complicated multi-orbital Fermi surface and a large $p$-$d$ hybridization which distributes the spectral weight of the bands over a wide energy range, susceptibility calculations are extremely delicate. In particular, 
one should avoid the CMA, as it can lead to wrong results, and carefully monitor the convergence of the results with the number of bands. The latter \textit{caveat} is particularly relevant for model studies of trends in FeSCs based on downfolded models of the electronic
structure~\cite{kuroki_pnictogen_2009,kemper_sensitivity_2010}.

Needless to say, the convergence of the interacting susceptibility with the number of bands might differ, since $s$, $p$, and $d$ bands will respond differently to correlations due to different interaction parameters.

\section{Dynamic Bare Susceptibility}
\label{sec:dynamic_susceptibility}

\begin{figure}
  \begin{center}
    \includegraphics[width=1.0\linewidth]{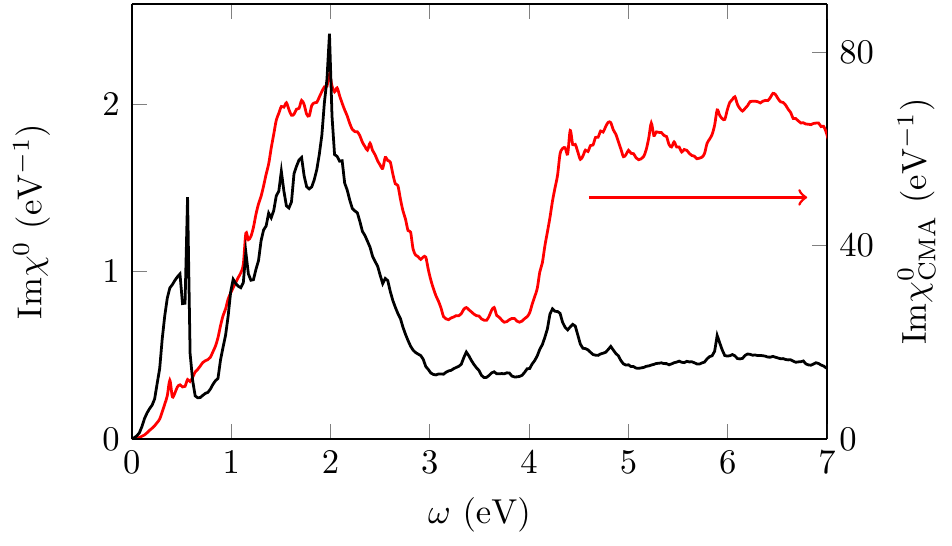}
  \caption{(Color online) Imaginary part of the dynamic bare susceptibility $\chi^0$ for FeSe halfway between $\Gamma$ and M. The black line represents the result obtained with exact matrix elements and the red line depicts the CMA result.}
  \label{fig:Imchi0_FeSe_0_MGhalf}
  \end{center}
\end{figure}

\begin{figure*}
  \begin{center}
    \includegraphics[width=1.0\linewidth]{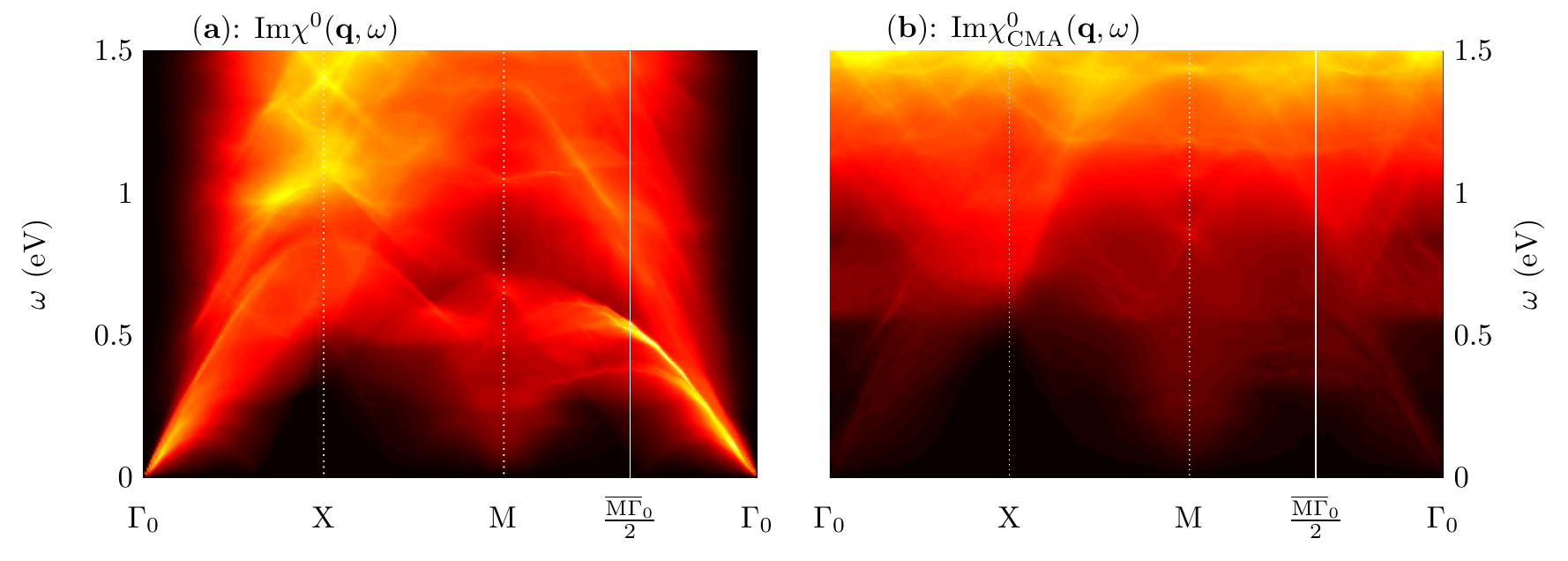}
  \caption{(Color online) Imaginary part of the dynamic bare susceptibility for FeSe at ambient pressure. The left panel ($\mathbf{a}$) shows $\text{Im}\chi^0$ whereas the right panel ($\mathbf{a}$) contains the results for $\text{Im}\chi^0_{\text{CMA}}$. The $\mathbf{q}$ point shown in \fig{fig:Imchi0_FeSe_0_MGhalf} is indicated by a vertical white line.}
  \label{fig:Imchi0_FeSe_4_surf_big}
  \end{center}
\end{figure*}

In addition to the static susceptibility, which is connected to
instabilities towards ordered ground states, valuable information
can be obtained also from the {\em dynamic} susceptibility. This
quantitity describes the elementary excitations of the system. We
focus here on its
imaginary part, which is directly related to scattering experiments
and has therefore a transparent physical interpretation. We show
below that the approximations discussed in the previous section for
the static susceptibility have even more dramatic effects in the
dynamical case. Of course, direct comparison to experiments requires
knowledge of the full interacting susceptibility, which is beyond the
scope of this work. However, a crucial ingredient to all theoretical
descriptions is a proper calculation of the bare susceptibility, which
we discuss here.

In \fig{fig:Imchi0_FeSe_0_MGhalf} we show the frequency dependence of
$\text{Im}\chi^0$ for a representative $\mathbf{q}$ point in the \BZ,
which sits half-way between $\Gamma$ and M. The black line shows the
result with all matrix elements properly included. As compared to the
CMA result, the matrix elements
strongly enhance some parts of the spectrum and suppress others. For
example, the small shoulder around $0.5$\,eV in the CMA result (red
line) is enhanced forming a well-defined peak, while the high-energy
contributions are strongly suppressed.
The reason for this deviation is again that the CMA completely
neglects the effect of orbital character, leading to an
overestimation of certain transitions. This is most obvious for
energies above 4\,eV, where the discrepancy increases sharply. At this
energy, the transitions are to a very large extent from the
  bands with dominant Se~$p$ orbital character to those with dominant
  Fe~$d$ and vice versa. In this case, the matrix elements are small
compared to direct $d$-$d$ transitions; this effect is not at all
reflected in the red curve of \fig{fig:Imchi0_FeSe_0_MGhalf}. As a
result, the overall spectral weight in the CMA is too large at high
frequencies. The susceptibility even shows a linear increase
for very large frequencies, which makes the use of a Kramers-Kronig
transformation meaningless. 

Figure~\ref{fig:Imchi0_FeSe_4_surf_big} shows intensity plots along high
symmetry lines in the \BZ \ for $\text{Im}\chi^0$ (left) and
$\text{Im}\chi^0_{\text{CMA}}$ (right). The $\mathbf{q}$ point used in
\fig{fig:Imchi0_FeSe_0_MGhalf} is indicated by a vertical white
line. The narrow peak at approximately $0.5$\,eV in
\fig{fig:Imchi0_FeSe_0_MGhalf} translates into a well-defined branch
of single-particle excitations, extending up to $0.7$\,eV. Another
high-intensity region of $\text{Im}\chi^0$ starts around $1$\,eV,
concentrated at the X point. In the CMA results, the spectral weight
distribution is very different. For example, the low energy branch is
almost completely suppressed and a large, featureless continuum above
$1.5$\,eV appears. 

The bare spectrum as presented here contains information about
the single-particle excitations of the system, and these can be
measured by inelastic neutron scattering. However, their intensity is
rather weak compared to collective excitations, such as
(para)magnons. 

As mentioned above, these require a calculation of the
full susceptibility, which is highly non-trivial. 
Even if one of the simplest approximations is used, 
namely the random-phase approximation (RPA)~\cite{hirschfeld_gap_2011,chubukov_magnetism_2008}, further assumptions on the interaction Hamiltonian are needed to make the calculation feasible;
in particular, the computational cost grows with the number of included bands, and this requires downfolding
the electronic structure to an effective low-energy model. 
More refined methods exist - FLEX, fRG - which improve the treatment of many-body interactions, 
but they are even more expensive computationally.
An
alternative approach that treats the interacting kernel
\textit{ab initio} has been suggested recently by Essenberger~\etal~\cite{essenberger_paramagnons_2012}.

However, every calculation for the full susceptibility relies on
an accurate evaluation of the bare susceptibility.
Following results of
Ref.~\onlinecite{essenberger_paramagnons_2012}, we want to note that
the position of collective excitations is crucially
influenced by the precise structure of the bare susceptibility. 
For instance, (para)magnon dispersions form in regions of the
$\mathbf{q}$-$\omega$ plane, where the intensity of the
single-particle excitations is low. 
This of course means that a
precise calculation of the bare susceptibility, as we present it in
this work, is an absolutely necessary ingredient also
for an accurate calculation of the interacting susceptibility, which
can then be compared to experimental results.

\section{Conclusions}
\label{sec:conclusions}

In this paper we have presented a practical implementation of bare static and dynamic susceptibilities, based on full-potential LAPW calculations, and a very efficient tetrahedron method for $k$-space integration. This allowed us to study in detail the effect of matrix elements and the convergence with the number of bands for some representative and topical materials (Cr, FeSe). We were able to show that the approximation, where all matrix elements in the susceptibility formula are replaced by a constant value (CMA), can lead to unreliable results. Therefore, nesting arguments, which are based on this approximation and are often employed to explain instabilities towards different orderings, are many times unfounded.~\cite{borisenko_two_2009,zhang_itinerant_2010}

Moreover, we have studied the convergence of the results as far as the summation over the bands is concerned. It appears that the convergence is slower then commonly expected. This could affect schemes that are based on downfolded models of the electronic structure, where only a few bands are taken into account. 
These effects, already significant in the static susceptibility, are even more severe for the dynamic susceptibility. For instance, low-energy excitations might not be visible or misplaced when matrix elements are not treated properly.

In view of the unpredictable accuracy  of CMA and/or a band-summation restriction it is advisable to refrain from any approximations and to evaluate the susceptibility formula exactly. One possible and efficient approach has been presented in this paper.

\acknowledgments{We are grateful to E. Schachinger for insightful discussions. Calculations have been done on the dcluster of TU Graz. L.B. acknowledges financial support from DFG, Project No. SPP1458 BOE3536-2. We thank NAWI Graz for support.}

%

\end{document}